\documentclass[sigconf]{acmart}

\AtBeginDocument{%
  }

\setcopyright{acmlicensed}
\copyrightyear{2024}
\acmYear{2024}
\acmDOI{XXXXXXX.XXXXXXX}

\acmConference[Conference acronym 'XX]{Make sure to enter the correct
  conference title from your rights confirmation emai}{June 03--05,
  2018}{Woodstock, NY}
\acmISBN{978-1-4503-XXXX-X/18/06}

\usepackage{multirow}
\usepackage{array}
\begin{document}

\title{From Perception to Decision: Assessing the Role of Chart Types Affordances in High-Level Decision Tasks}


\author{Yixuan Li}
\affiliation{%
  \institution{Georgia Institute of Technology}
  \city{Atlanta, GA}
  \country{USA}}
\email{yixuanli@gatech.edu}

\author{Emery D. Berger}
\affiliation{%
  \institution{University of Massachusetts Amherst}
  \city{Amherst, MA}
  \country{USA}
}
\email{emery@cs.umass.edu}

\author{Minsuk Kahng}
\affiliation{%
 \institution{Yonsei University}
 \city{Seoul}
 \country{Republic of Korea}}
\email{minsuk@yonsei.ac.kr}

\author{Cindy Xiong Bearfield}
\affiliation{%
  \institution{Georgia Institute of Technology}
  \city{Atlanta, GA}
  \country{USA}}
\email{cxiong@gatech.edu}

\renewcommand{\shortauthors}{Yixuan et al.}

\begin{abstract}
Visualization design influences how people perceive data patterns, yet most research focuses on low-level analytic tasks, such as finding correlations.
 The extent to which these perceptual affordances translate to high-level decision-making in the real world remains underexplored. 
 Through a case study of academic mentorship selection using bar charts and pie charts, we investigated whether chart types differentially influence how students evaluate faculty research profiles. 
 Our crowdsourced experiment revealed only minimal differences in decision outcomes between chart types, suggesting that perceptual affordances established in controlled analytical tasks may not directly translate to high-level decision scenarios. 
 These findings emphasize the importance of evaluating visualizations within real-world contexts and highlight the need to distinguish between perceptual and decision affordances when developing visualization guidelines.
\end{abstract}




\maketitle
\section{Introduction}

Visualizations transform data into insights and decisions, shaping how we act upon information. 
Visualization design can afford different reader perceptions of data~\cite{bearfield2024same, holder2023polarizing, liu2025binary}.  
Existing research has evaluated visualization affordances primarily through analytic tasks, such as identifying correlations~\cite{harrison2014ranking, kim2018assessing} and comparing values~\cite{fygenson2023arrangement}. 
These evaluations have generated perception-based guidelines such as using bar charts to facilitate data magnitude comparisons~\cite{xiong2022reasoning} and pie charts to emphasize part-to-whole relationships~\cite{kosara2019evidence, eells1926relative}. 

However, in the real world, visualizations serve as scaffolding for complex decision-making beyond basic perception tasks.
Whether the perceptual benefits of these guidelines directly carry over to real-world decision-making contexts remains an open question.
Our work addresses this gap by investigating whether chart type affordances can significantly impact data-driven decision outcomes when applied to real-world situations, using pie charts and bar charts as our case study. 



We situate our case study within the context of students selecting research mentors in universities, which is a particularly relevant scenario given the increasing trend of students pursuing graduate education~\cite{duncan2020enrollment}.
In this setting, students are making high-level decisions by evaluating faculty members' research productivity and interdisciplinary engagement. 
Popular platforms such as CSRankings~\cite{csranks} often present this information through data visualizations, as shown in Figure~\ref{fig:4Profiles}. 
These visualizations can affect students' perceptions of faculty publication profiles and interdisciplinary scope, potentially influencing their decisions.

Bar charts, which facilitate magnitude comparisons, may emphasize publication counts overall, potentially focusing the student on a faculty member's total research output. 
Conversely, pie charts, which highlight proportional relationships, may accentuate disciplinary categories, potentially focusing the student on the interdisciplinarity of the faculty's research program. 
If perceptual affordances influence high-level decisions, we would expect to see differential effects in students' decisions depending on the chart type.

\vspace{2mm}
\noindent \textbf{Contribution and Takeaways:}
We compare the effect of chart types on high-level decisions through a crowdsourced experiment. 
By surveying visitors to a popular metrics-based computer science institution ranking website CSRankings.org~\cite{csranks}, we examined whether visualization choice significantly affects impressions of faculty research profiles. 
Our findings suggest a small effect in decision outcomes between bar charts and pie charts, suggesting that the perceptual affordances of these chart types did not substantially translate to high-level decision making. 
These results emphasize the importance of evaluating visualization affordances in high-level decision contexts in addition to traditional low-level tasks.
We also hope to encourage designers and researchers to distinguish between perceptual affordances and decision affordances when developing visualization guidelines.

\section{Related Work}

Visualization design can shape data interpretations. 
Existing work has extensively explored the affordance of chart types in visual data communication, with the effectiveness of pie charts being one of the most contentious topics in the literature~\cite{evergreen2019effective, skau2016arcs}.
The visualization community has condemned pie charts for their ineffectiveness for many tasks~\cite{midway2020principles, siirtola2019cost}.
Pie charts are notoriously challenging for visual comparison tasks~\cite{annesley2010bars}.
While a visualization reader can rely on area-comparison to make sense of data values in bar charts and pie charts, bar charts offer position as another encoding channel readers can rely on, where pie charts offer angle and arc length~\cite{cleveland1984graphical, mackinlay1986automating}.
Humans perception of area comparison is imprecise \cite{munzner2014visualization}, and angle is one of the least important visual cues~\cite{skau2016arcs}.
Furthermore, comparing the length of visual marks is much more straightforward when the length is encoded via a straight line as in bar charts, compared to the arc in pie charts. 
Therefore, pie charts can be ineffective for many visualization tasks from a perceptual perspective. 
For example, for a range of low-level visualization tasks~\cite{amar2005low}, such as finding clusters and finding anomalies, pie charts lead to significantly less accurate and lower performance~\cite{saket2018task}.


However, pie charts are not always as ineffective as they are often portrayed by the visualization community.
Pie charts can be especially useful in communicating part-to-whole relationships and facilitating proportional judgments~\cite{eells1926relative, spence1991displaying}, even more so than bar charts~\cite{eells1926relative}. 
When the number of data points visualized is small, people can perform data estimation tasks faster and more accurately with pie charts compared to bar charts~\cite{eells1926relative}.
Further, when visualizing categorical data (with up to five categories) using pie charts versus divided bar charts, pie charts are associated with higher performance~\cite{croxton1927bar} (although work by Simkin et al.~\cite{simkin1987information} and Spence et al. ~\cite{spence1991displaying} have showed bar and pie charts to lead to comparable performance levels).  

These prior investigations into visualization affordances have focused on perception-based tasks where researchers evaluate chart types through metrics like task speed, accuracy, and user preferences~\cite{sedlmair2016beliv}. 
This leaves a gap in understanding how visualizations influence readers in more complex decision-making contexts. 
These higher-level decisions go beyond simply extracting data values and require the user to interpret information in a broader context~\cite{burns2020evaluate}. 
To make informed inferences or predictions, visualization readers must synthesize the presented information~\cite{burns2021designing}. 
There has only been limited work in this space.
One study by Holder et al.~\cite{holder2022dispersion}, for example, studied the effect of stereotyping with chart types.
They found that, compared to jittered dot plots, bar charts can make readers form stereotypical impressions of groups. 
Our work addresses this gap by exploring visualization affordances in higher-level decision-making, using decision-making in higher education as a case study.

\section{The Study}
We compared the effect of chart type on high-level decision-making in the context of academic decisions.
We presented synthesized faculty profiles either as pie charts or bar charts, and surveyed people's impression of their productivity and interdisciplinarity, and ultimately their willingness to be advised by these faculty members. 


\begin{figure}
  \centering
  \includegraphics[width=1\linewidth]{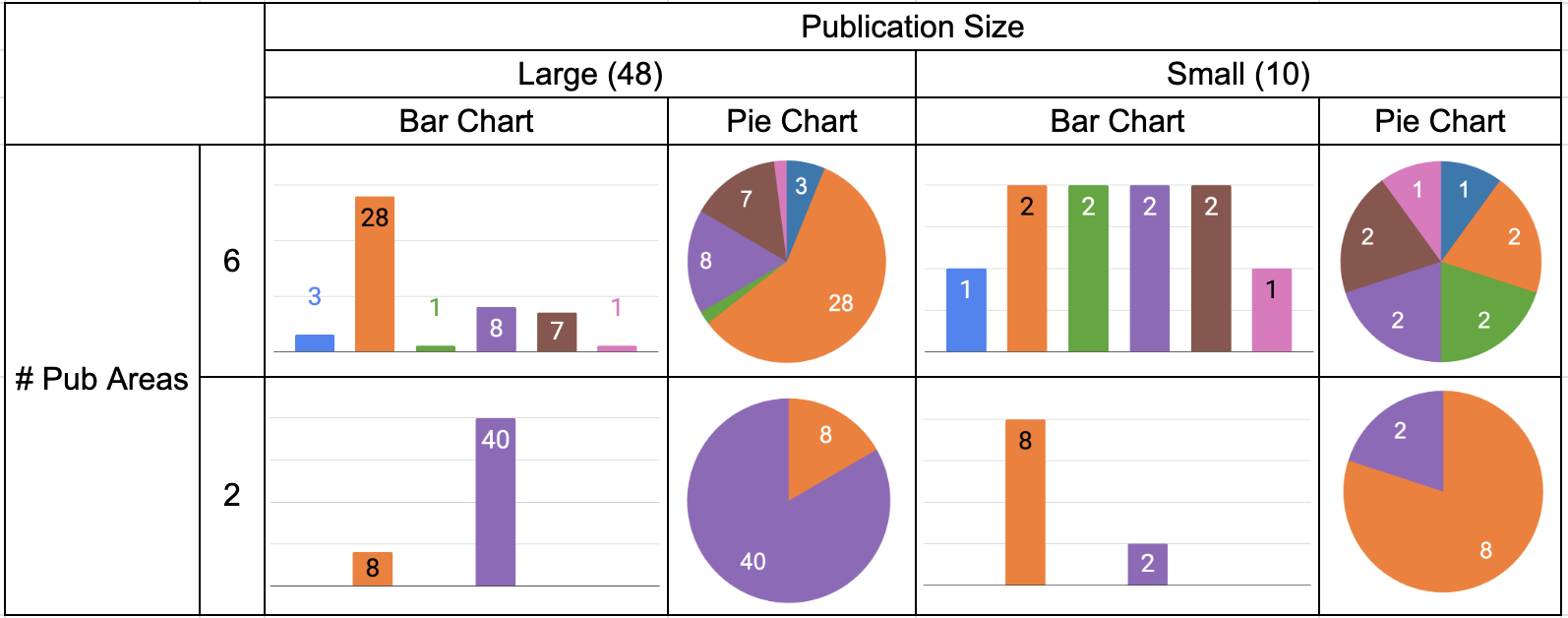} \centering
  \caption{Bar Chart and Pie Chart Stimuli for the four conditions: 6L (faculty with Large publication size in 6 areas), 2L (faculty with Large publication size in 2 areas), 6S (faculty with Small publication size in 6 areas), 2S (faculty with Small publication size in 2 areas).}
  \label{fig:4Profiles}
\end{figure}

\subsection{Stimuli}

To ensure the real-world relevance of our findings, we linked our survey to the computer science academic ranking website CSRankings.org~\cite{csranks}.
The website is widely used by students and faculty, with web traffic analysis revealing that it had more than 70,000 visits from March to April 2025.

\vspace{2mm}
\noindent\textbf{Survey Distribution: }
For the short duration when the survey was made available on CSRankings, visitors can choose to view faculty profiles via one of these two chart types (pie or bar) to see their publication records across research areas. 
Prospective students can use these charts to identify 
Institutions and faculty members for pursuing higher degrees in computer science.

\vspace{2mm}
\noindent\textbf{Visualization Design: }
We created pie charts and bar charts to visualize synthesized faculty publication profiles across research areas, referencing data from CSRankings. 
Our observations revealed that faculty profiles tend to vary along two dimensions: publication volume (many vs. few) and disciplinary breadth (many vs. few areas). 
Based on these patterns, we designed four profile conditions to reflect these dimensions at a high level, as illustrated in Figure~\ref{fig:4Profiles}.
We varied the range of discipline areas the faculty has published in to be either very interdisciplinary (6 areas) or very focused (2 areas). 
We varied the number of publications to either be large (L-48 papers published) or small (S-10 papers published).
We refer to these conditions as 6L, 6S, 2L, and 2S. 

\begin{table*}[t]
  \caption{Types of questions used in the study to assess participants' impressions of faculty profiles.}
  \label{Tab:question}
  \centering
  \begin{tabular}{p{1cm}p{3.5cm}p{10cm}p{1.5cm}}
   \hline
  \textbf{Factor} & \textbf{Type} & \textbf{Question} & \textbf{Mean (SD)} \\
  \hline
  \raggedright   & \raggedright Willingness to work \arraybackslash &  How much would you want to work with them? & 4.14 (1.33)\\
  \hline
\multirow{3}{*}{\raggedright F1} & \raggedright Research Interdisciplinarity \arraybackslash & How interdisciplinary do you think the faculty member's research program is? &3.66 (1.41)\\
& \raggedright Cross-Area Collaboration \arraybackslash & In addition to your advisor, how often do you expect to work with faculty members outside your primary research area? & 3.28 (1.33)\\
& \raggedright Meeting Frequency \arraybackslash & How often do you think you will get 1-on-1 meetings with your advisor?& 3.62 (1.09)\\
\hline
\multirow{3}{*}{\raggedright F2} & \raggedright Faculty Productivity \arraybackslash & How productive do you think the faculty member is? & 4.16 (1.43)\\
& \raggedright Workload Expectation \arraybackslash & How busy do you expect to be while working with your advisor? &4.04 (1.21)\\
 & \raggedright Group-Size Expectation \arraybackslash & How small/large do you expect your advisor's research group to be? &3.43 (1.61)\\
   \hline
  \end{tabular}
  \label{questionTable}
\end{table*}

\subsection{Procedure}

During our study, visitors to CSRankings.org~\cite{csranks} had the option to click on the survey link and complete our survey via Qualtrics~\cite{qualtrics}.
Upon consenting to the study, participants answered two questions about the importance of faculty publication volume and disciplinary breadth in shaping their advisor preferences, to account for their prior beliefs.
After that, participants were randomly assigned to see one of the four research profiles (6L, 2L, 6S, and 2S) as a pie chart or a bar chart. 
We asked participants to answer seven Likert-scale questions from 0 (Not At All) to 6 (Very) 
of their \textbf{impression} of faculty research productivity, interdisciplinarity, projected workload, which existing work has suggested to impact one's general willingness to work with a faculty member~\cite{mishra2021choice, roy2023doctoral}.
See Table~\ref{Tab:question} for the actual questions and basic statistics.


\subsection{Participants}
We crowdsourced data via a volunteer link on CSRankings.org, shown to visitors through a randomized overlay. In total, we received 1,053 responses, with 212 complete after filtering incomplete surveys.

\subsection{Hypothesis}
\begin{figure}
  \centering
  \includegraphics[width=1\linewidth]{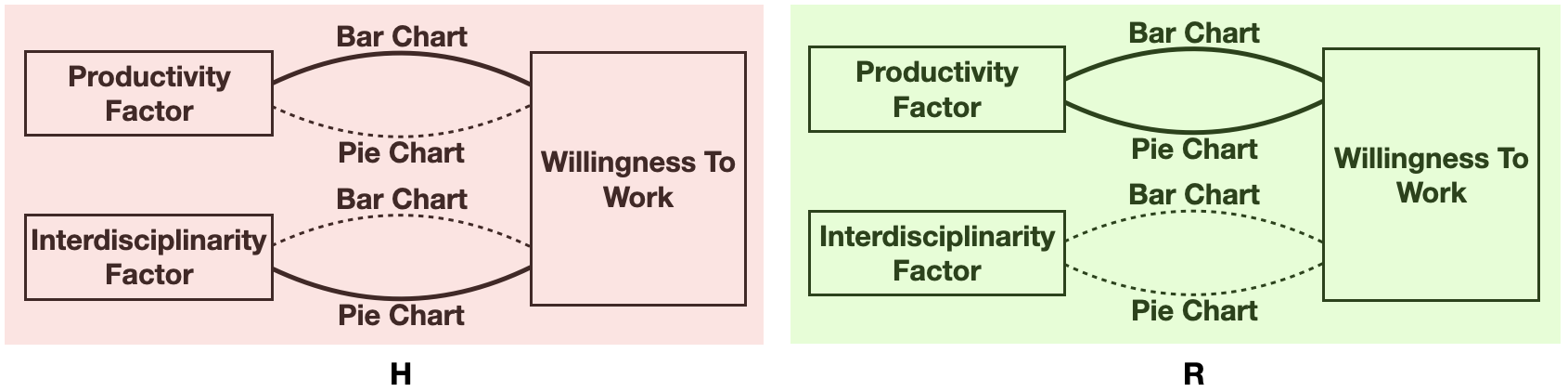} \centering
  \caption{The hypothesized model (H), where Bar chart leads to stronger correlation for Productivity Factor, while Pie chart leads to stronger correlation for Interdisciplinarity Factor, and the actual result (R), while Productivity always has a stronger correlation with Willingness to Work, regardless of chart types.}
  \label{fig:4Results}
\end{figure}

As illustrated in Figure \ref{fig:4Results}, we hypothesized that chart type affordances would extend to high-level decision tasks, with visualization format systematically influencing evaluations of faculty profiles. Specifically, we predicted that bar charts would emphasize magnitude information (publication count),
while pie charts would highlight categorical distribution (disciplinary breadth). 
Consequently, participants would demonstrate different willingness to work with faculties depending on whether the profile visualization emphasized productivity (bar charts) or interdisciplinarity (pie charts). 
\section{Result}


We asked participants six questions, ranging from meeting frequency to workload expectations (see Table~\ref{questionTable}), that may influence a student’s willingness to work with a faculty member. 
A linear regression analysis revealed that these six factors significantly predicted willingness to work with the faculty member (Adjusted $R^{2} = 0.457$, $F(6, 205)$ = $30.6$, p $< .001$).

We then conducted a factor analysis on participants’ responses to assess the latent variables underlying their evaluations of faculty profiles. 
The analysis revealed two orthogonal factors that accounted for the majority of variance: interdisciplinary breadth (Factor 1) and research productivity (Factor 2).
These factors reflect distinct constructs that guide participants' willingness to work with a faculty member. 

To test our hypothesis regarding the affordance of visualization design on decisions, we computed the correlation between each factor score and participants’ reported willingness to work with the faculty member under two visualization conditions (Bar Chart vs. Pie Chart). We further quantified the magnitude of distributional differences in willingness scores across conditions using Earth Mover’s Distance (EMD) as a measure of effect size. 
Results suggests a minor effect of visualization design on decisions. 
All statistical analyses were conducted using R and details can be found in the supplementary materials. 

\subsection{Factor Analysis}
\begin{figure}
  \centering
  \includegraphics[width=1\linewidth]{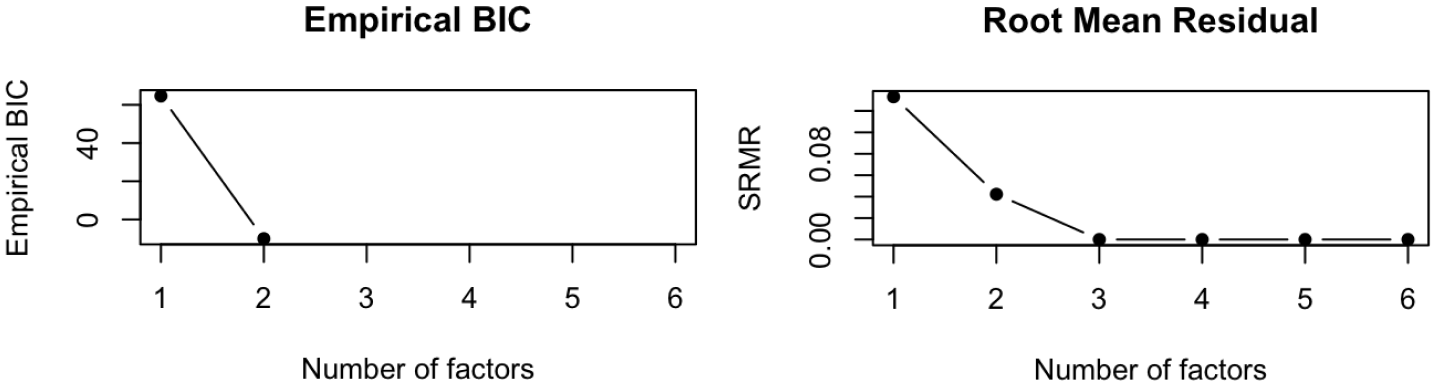} \centering
  \caption{The empirical BIC and root mean residuals of factor models consisting of 1 to 6 factors.}
  \label{fig:exp1_factorAnalysis}
\end{figure}

We first conducted a factor analysis using the \texttt{Psych} \texttt{R} package \cite{revelle2015package} to identify factors that underly participants' evaluations of faculty profiles. 
We compared the empirical BIC and root mean residuals of factor models consisting of 1 to 6 factors, as shown in Figure ~\ref{fig:exp1_factorAnalysis}. 
We found that the empirical BIC reached the lowest at two factors. 
Root mean residuals decreased linearly with added factors, with diminishing returns after two.
Based on these metrics, we decided to apply the two-factor model to our survey: 
\textbf{Factor1-Interdisciplinarity} is about \emph{how interdisciplinary the faculty member is}, including questions of research interdisciplinarity, cross-area collaboration, and meeting frequency. 
\textbf{Factor2-Productivity} is about \emph{how productive the faculty member is}, including questions of faculty productivity, workload expectation, and group-size expectation.
Interestingly, these two factors correspond to the distinct perceptual affordances emphasized by bar charts (e.g., supporting magnitude comparison) and pie charts (e.g., highlight part-whole relationships). 
Based on this alignment, we use these factors in subsequent analyses to test our hypothesis.



\subsection{Chart Types Impact on Impression}

To further investigate how chart type might afford impressions associated with each factor, we conducted a correlation coefficient comparison between these factors and willingness to work under each chart type, as shown in Table \ref{Tab:cor}. This allowed us to assess whether one chart type made participants weight a given factor more heavily in their decisions.
Fisher’s z-test \cite{fisher1970statistical} and Zou’s confidence intervals \cite{zou2007toward} were performed using the \texttt{cocor} \texttt{R} package \cite{diedenhofen2016package} to assess whether the differences in correlation coefficients were significant.
We found no significant difference between bar and pie charts in the correlation of either F1 (d = $0.047$, z = $0.338$, p $> 0.05$) or F2 (d = $0.148$, z = $1.515$, p $> 0.05$) with willingness to work.
This suggests that while there are slight variations in how bar charts and pie charts influence participants' focus on productivity and interdisciplinarity, 
both chart types lead to comparable decision-making outcomes in this context.

We also tested whether the strength of the association between willingness to work and either productivity (F2) or interdisciplinarity (F1) varied by chart type, using Hittner, May, and Silver’s modified z-test~\cite{silver2004testing} and Zou’s confidence intervals~\cite{zou2007toward}. 
For bar charts, F2 showed a significantly stronger correlation with willingness than F1 (d = $-0.457$, z = $-4.519$, p $< .001$). 
Pie charts showed a similar pattern (d = $-0.355$, z = $-2.917$, p $< .001$). 
In both cases, Zou’s confidence intervals excluded zero, suggesting statistical significance. 
These results suggest that across both chart types, participants weighted productivity (F2) more heavily than interdisciplinarity (F1) in their decision-making.



These results do not support our hypothesis, suggesting that chart type affordances do not significantly influence high-level decision-making. 
However, the effect of chart type was also overshadowed by the main effect of research productivity, which dominated participants' perception on willingness to work with a faculty. 
To further isolate visualization design-driven affordances, we conducted an effect size comparison via Earth Mover's Distance to assess how responses differed between the bar chart and pie chart conditions.


\begin{table}[t]
  \caption{Correlation between factors and willingness to work.}
  \label{Tab:cor}
  \centering
  \begin{tabular}{p{3cm}p{2cm}p{2cm}}
   \hline
  \textbf{Correlation} & \textbf{Bar Chart} & \textbf{Pie Chart} \\
  \hline
  cor (F1, Willingness) & 0.158& 0.111\\
  cor (F2, Willingness) & 0.615& 0.467\\

   \hline
  \end{tabular}
\end{table}

\subsection{Chart Type Effects via Distribution Comparison}
\begin{figure}
  \centering
  \includegraphics[width=1\linewidth]{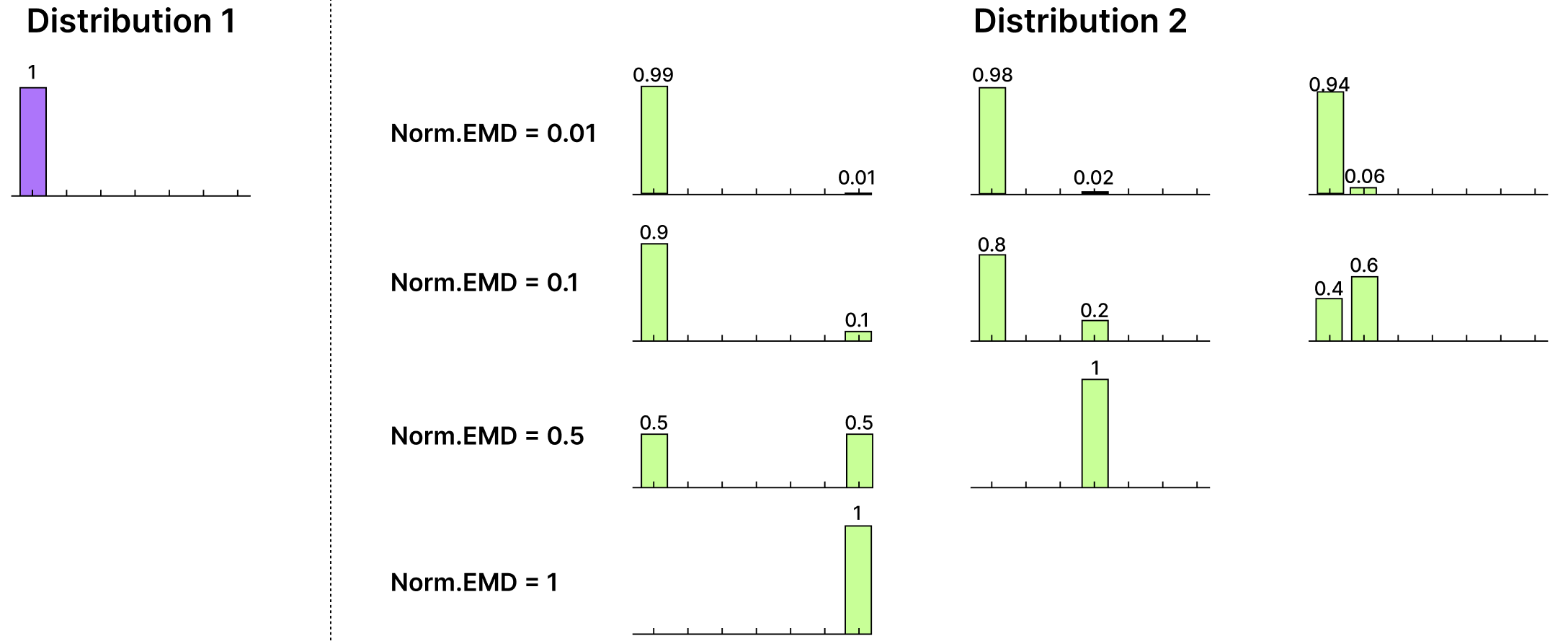} \centering
  \caption{An illustration demonstrating that when the normalized Earth Mover's Distance value (Norm.EMD) is equal to 0.01, 0.1, 0.5, and 1, with a fixed Distribution 1 on the left side, what Distribution 2 would look like on the right side. }
  \label{fig:NormEMD}
\end{figure}

\begin{figure*}[t]
  \centering
  \includegraphics[width=\textwidth]{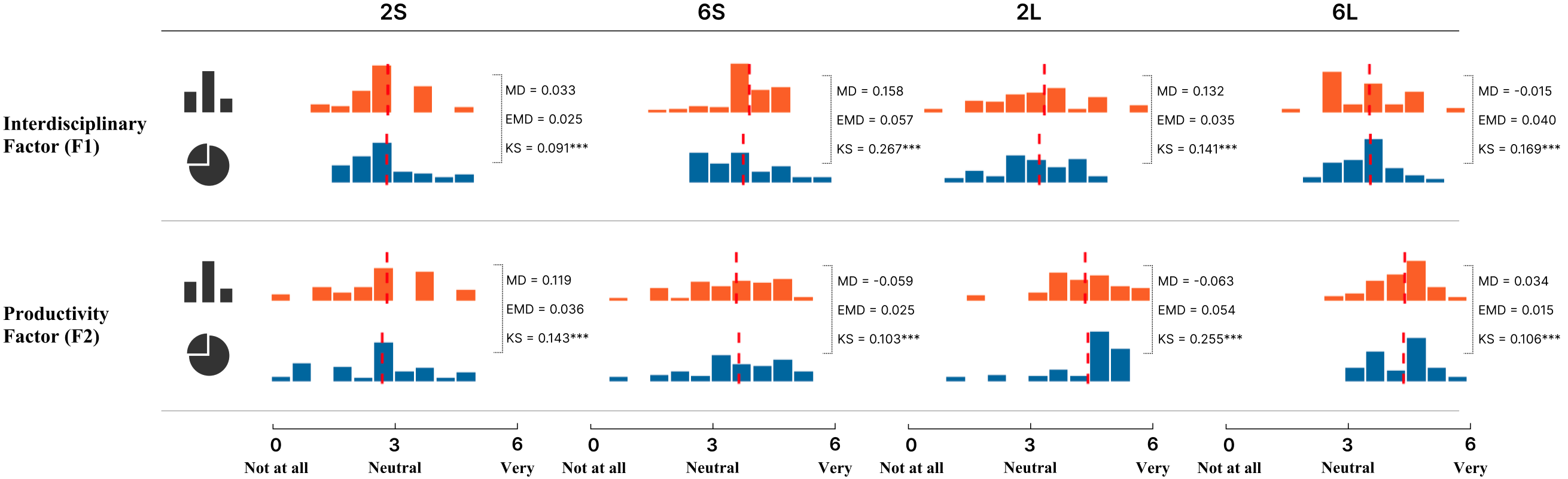} \centering
  \caption{Comparison of participant responses using bar charts (orange) and pie charts (blue) across four conditions (6L, 2L, 6S, 2S) for the two factors, with the red dash-line representing the average value of these factors. The diagrams show the Mean Difference (MD), normalized Earth Mover's Distance (Norm.EMD), and KS statistics, highlighting the impact of chart type on participant perceptions for each condition.* represents the significance of the KS test, with * representing $p-value<0.05$, ** representing $p-value<0.01$, and *** representing $p-value<0.001$. }
  \label{fig:result}
\end{figure*}

We calculated the Earth Mover's Distance (EMD)~\cite{rubner2000earth} between bar and pie charts to compare their effects in shaping high-level decisions regarding faculty impressions. 
EMD represents the minimal cost that must be paid to transform one distribution into the other, which is a robust metric capturing the overall distribution `shape' features instead of just the maximum difference between the cumulative distribution functions, such as Kolmogorov-Smirnov (KS) test \cite{massey1951kolmogorov}, or just the difference between means, such as t-test. 
Calculating the EMD values between the bar and pie charts helps us understand the effect size of chart types on participants' responses. 

However, one limitation of EMD is that there is no maximum bound or p-value for statistical significance comparison. 
To address this issue, we computed the normalized EMD value (Norm.EMD) using the following equation: 
\begin{equation}
   Norm.EMD = \frac{EMD}{Max\ Possible\  EMD}
\end{equation}
where Max Possible EMD is computed by the greatest distance that the `mass' of one distribution would need to be moved to match the other distribution. 
Here, normalized EMD ranges from 0 to 1, where 0 means that the two distributions are identical and 1 means that the two distributions are entirely non-overlapping. 
Since our survey required participants to respond on a seven-point scale, we calculated EMD based on histograms with seven bins.
Figure~\ref{fig:NormEMD} illustrates example effect sizes of the normalized EMD between Distribution 1 and various versions of Distribution 2, providing an interpretive reference for understanding EMD values in our results.

We calculated the normalized EMD 
between the distributions of responses for pie and bar charts across all four conditions and both factors, based on 1,000 bootstrap samples (see Figure~\ref{fig:result}).
The resulting normalized EMD values ranged from 0.019 to 0.057, indicating that the overall distributional differences between chart types were relatively small across all comparison groups.
Notably, even in cases where the mean difference (MD) between chart types was relatively large, such as in the 6S condition, where the MD for the F1 reached 0.158, the corresponding normalized EMD remained modest (0.057). 
This suggests that shifts in distribution means were not accompanied by substantial changes in the overall distribution.

We also computed the Kolmogorov–Smirnov (KS) statistics \cite{massey1951kolmogorov} to provide an additional point of comparison. Unlike the normalized EMD, which considers the overall 'effort' to transform one distribution into another, the KS statistic captures the maximum difference between the cumulative distribution functions (CDFs) of the two chart types using a non-parametric approach. 
We found the KS differences ranged from 0.091 to 0.267 across all comparisons.
While these are mostly statistically significant, the effect size is small, suggesting only a small effect of chart types. 
There are cases where the normalized EMD is small but the KS statistic is large. 
For instance, in the 6S condition for Interdisciplinarity Factor (F1), the KS statistic is 0.267, suggesting a moderate difference between the two cumulative distribution functions (CDFs) (see Figure~\ref{fig:result} for 6S, F1).
This result shows that participants gave different ratings of interdisciplinary breadth to faculty positions when looking at bar charts compared to pie charts, but the difference remains small (Norm.EMD = 0.057). 
Together, these findings suggest that while chart type can influence how participants respond to questions related to productivity and interdisciplinarity, the magnitude of the effect is generally small. 

\section{Discussion}
We reflect on our findings and discuss their implications for visualization design and future research.

\subsection{Implications for Visualization Affordances}

Our findings suggest that while chart type can significantly shape user perceptions in low-level visual analysis tasks~\cite{kim2018assessing, saket2018task}, such perceptual affordances may not significantly influence high-level decision-making.
Although bar and pie charts differ in their perceptual affordances, in that bar charts emphasize magnitude comparisons and pie charts emphasize proportion comparisons, these differences did not translate into meaningful differences in high-level decisions in our case study.

Together, these results highlight the need to study visualization affordances in high-level decision-making contexts in real-world scenarios. 
This finding presents an opportunity to develop a framework that connects visualization affordances at the perceptual level to higher-level decision-making in real-world contexts.

\subsection{Limitations and Future Direction}

We identify several limitations in this study that open up opportunities for future research.

\textbf{Chart Types:} While we found that using bar charts or pie charts might not significantly impact viewers' impressions of faculty profiles, our comparison did not extend to other chart types not used on CSrankings.org. 
Future work should explore whether other chart types or data representation choices may have different affordances, for a different context or decision-making task. 

\textbf{Faculty Profile Scope: }The stimuli only represented publication count and disciplinary range, consistent with the information displayed on CSrankings.org. However, there may be other relevant aspects of faculty profiles, such as the number of graduate or undergraduate students in their lab, that could be influenced by the choice of chart type. 
Future work can explore additional data dimensions in visualizations.

\textbf{Cognitive Task Types: }The questions we asked participants ranged from general willingness to work with the faculty to the frequency of personal one-on-one meetings. These can be further categorized based on task type, following existing cognitive task taxonomies, such as Dimara's Taxonomy of Cognitive Biases for Information Visualization~\cite{dimara2018task}. Future work could systematically examine how chart design affects specific task types using cognitive bias taxonomies.

\textbf{Participant Demographics:} 
According to Google Analytics records~\cite{googleAnalyticsTools}, only 7.9\% of the 53K visitors to the study site (August 15–31, 2024) were identified as female. While this may reflect limited disclosure of demographic information by site visitors, it also suggests that our participant pool may have been heavily gender-imbalanced. 
Although our study did not focus on gender differences in decision-making, further research is needed to assess whether the findings generalize across demographic groups.


\bibliographystyle{ACM-Reference-Format}
\bibliography{sample-base}


\end{document}